\documentclass[twocolumn,showpacs,preprintnumbers,amsmath,amssymb]{revtex4}
\usepackage{graphicx}
\usepackage{dcolumn}
\usepackage{bm}

\begin{document}


\title{ On the diffusion coefficient of a photon migrating through a turbid medium: a fresh look from a broader perspective}

\author{K. Razi Naqvi}
\affiliation{Department of Physics,
Norwegian University of Science and Technology,
NO-7491 Trondheim, Norway}

\date{\today}

\begin{abstract}
Does the diffusion coefficient of a photon depend on time $t$ or the probability of absorption $k$? To find an answer to the question, photon transport in a medium of infinite extent is analyzed using the method of moments. It is pointed out that if $D$ is defined so as to make it depend on $t$ or $k$, it will also depend on the experimental conditions; that the parameter $k$ which enters the stationary diffusion equation is in general different from that entering the transient version; and that a hitherto unused non-Markovian partial differential equation may be used for treating photon transport.

\end{abstract}

\maketitle

A theoretical basis for quantifying translational diffusion in terms of a diffusion coefficient ($D$) was laid in 1905 by Einstein \cite{Eins:05, EinsFurt:56}; shortly afterwards, he published a second note \cite {Eins:06, EinsFurt:56}, wherein he acknowledged that his formula ``cannot be applied for any arbitrarily small time''. Einstein concluded that the diffusion equation (DE), which governs the density in coordinate space, as well as the definition of $D$ that follows from the equation ``only holds for intervals of time which are large compared with $\mu B$'';  the product $\mu B$, called the velocity relaxation time, will be denoted here by $1/\beta$ (see below). To go beyond the DE it becomes necessary to formulate a so-called transport equation for describing the density in phase space; it is not sufficient, nor indeed advisable (as shown below), to employ the telegraph equation (TE). The label Lorentz-Boltzmann equation (abbreviated as LBE) will be used here for a transport equation applied to photons and monoenergetic neutrons \cite{DudeMart:79}. Studies based on the LBE have given rise to a debate as to whether $D$ depends on time and/or the probability of absorption by traps in the host medium. A recent addition to this material is a series of papers by Alfano, Cai, Lax and Xu \cite{CLAX:02a, CLAX:02b, CLAX:05}, who concluded that $D$ depends on time but not on absorption; they inferred their definition of $D$ with the aid of an approximation based on a cumulant expansion of the distribution function $I({\bf r},{\bf c},t)$. Earlier, Aronson and Corngold \cite{AronCorn:99} had claimed the converse to be true, and cited some experimental evidence in support of their claim. My object here is to examine these issues (which arise also in other contexts) from a broader perspective.  

It will be convenient to begin by drawing attention to some implications of the cumulant approximation which appear to have been overlooked by its proponents, for whom I will use the alias CLAX (composed of their initials). CLAX proposed that the exact particle density $N({\bf r},t)$ in a system with plane symmetry (around the $z$-axis) can be replaced by a product of three Gaussians $N^{\rm (G)}({\bf r},t)\equiv G(x,t)G(y,t)F(z,t)$. Since $N^{\rm (G)}$ can be recovered by means other than cumulant expansion, I will refer to it as the Gaussian approximation (GAP, for short). My first task is to derive, using more elementary means and allowing for more general initial conditions, the expression for $F(z,t)$.

Let ${\bf e}_Z$ and ${\bf \Omega}$ be unit vectors along the $z$-axis and the velocity ${\bf c}$, respectively, $d{\bf \Omega}=d\mu d\phi$,  $\mu={\bf e}_Z\cdot {\bf \Omega}$, and define
\begin{equation} 
\Psi(z,\mu, t)\equiv 2\pi\int_{-\infty}^{\infty}\int_{-\infty}^{\infty}I({\bf r},\mu,t) dxdy,
\end{equation}
and
\begin{equation}
F(z,t)\equiv \int_{-\infty}^{\infty}\int_{-\infty}^{\infty}N({\bf r},t)dxdy=
\int_{-1}^{1}\Psi(z,\mu,t)d\mu.
\end{equation}

We will begin by assuming that absorption is absent and scattering is isotropic; with $\alpha =c/l=1/\tau$, where $l$ is the mean free path and $\tau$ the mean free time, the equation of radiative transfer can be written as
\begin{equation}
\Bigl [\partial_t +c\mu \partial_z \Bigr ]\Psi(z,\mu, t)=-\alpha \left [\Psi(z,\mu,t)-\textstyle{1\over 2}F(z,t)\right ].
\end{equation} 
Since the speed $c$ is taken to be a constant parameter, it will be convenient to call $\mu$ the velocity (along the $z$-direction).

{\it Calculation of moments.\/} We define the expectation of a quantity $\chi $ as $\overline{\chi}^{\mu_0}\equiv \int_{-1}^{1}d\mu\!\int_{-\infty}^{\infty}dz\,\chi \Psi(z,\mu,t;z_0,\mu_0)$, and proceed to calculate $\overline{z^2}^{\mu_0}$ for a photon with an initial velocity $\mu_0$ and initial position $z_0$. If the initial velocity distribution differs from $\delta (\mu-\mu_0)$, an ensemble average (denoted by an overline) can be found by taking a second average (over the initial velocities) of the first average: $\overline{\chi}=\overline{\overline{\chi}^{\mu_0}}$.  This notational device for distinguishing between the two averages is borrowed from Uhlenbeck and Ornstein (U\&O) \cite{UhleOrns:31}. For a medium of infinite extent we can insist that $F(z,t)\to 0$ as $|z|\to \infty$. If we multiply Eq.~(3) by $\chi=z^m\mu^n$ and integrate over all $z$ and all $\mu$, we obtain the equation
\begin{equation}
{d\over dt}\overline{\chi}^{\mu_0}-c\,\overline{\bigl[\mu {d\chi\over dz}\bigr ]}^{\mu_0}=
-\alpha \overline{\chi}^{\mu_0}+{\alpha\over 2}\!\int_{-\infty}^{\infty}\!\!\!dz F(z,t)\!\int_{-1}^{1}\!\!\!d\mu\, \chi,
\end{equation} 
which can be easily solved to get $d\overline{\chi}^{\mu_0}/dt$. For $(m,n)=(0,1)$, (1,0) and (0,2) one finds
\begin{eqnarray}
\overline{\mu}^{\mu_0}&=&\mu_0e^{-\alpha t},\\
\overline{z}^{\mu_0}&=&z_0+c\mu_0(1-e^{-\alpha t})/\alpha,\\
\overline{\mu^2}^{\mu_0}&=&\mu_0^2e^{-\alpha t}+\textstyle{1\over 3}(1-e^{-\alpha t}).
\end{eqnarray} 

The last relation is needed when one goes on to find and solve, by setting $(m,n)=(1,1)$, the equation satisfied by $\overline{z\mu}^{\mu_0}$. Finally, with $(m,n)=(2,0)$, the required equation for $\overline{z^2}^{\mu_0}$ comes out in the form
$d\overline{z^2}^{\mu_0}/dt=2c\,\overline{z\mu}^{\mu_0}$, which integrates to give
\begin{eqnarray} 
\overline{z^2}^{\mu_0} &=& z_0^2+2\tau \,c\mu_0z_0(1-e^{-\alpha t})+\textstyle{2\over 3}c^2\tau^2\left(\alpha t-1+e^{-\alpha t}\right )\nonumber\\
&&\quad\, +\textstyle{2\over 3}c^2\tau^2(3\mu_0^2-1)\,\left (1-e^{-\alpha t}-\alpha te^{-\alpha t} \right ).
\end{eqnarray} 
One can go on and find $\overline{z^n}^{\mu_0}$, for $n>2$; by availing oneself of symbolic computation, one can reduce the tedium and the risk of making errors. Once the raw moments are at hand, one can find the central moments $\overline{Z^n}^{\mu_0}$, where
\begin{equation} 
Z\equiv z-\overline{z}^{\mu_0}.
\end{equation} 
CLAX have allowed for anisotropic scattering; their expressions for the mean and the variance agree with mine if one sets $g_1=g_2$ in the former and $\mu_0=1$ in the latter; it will be sufficient to spell out the variance:
\begin{equation} 
\overline{Z^2}^{\mu_0=1}\!\!\!=\! \textstyle{1\over 3}c^2\tau\!\left[2t- (1-4e^{-\alpha t}+4\alpha te^{-\alpha t}
+3e^{-2\alpha t})/\alpha\right ].
\end{equation} 
For a Gaussian distribution, all odd central moments vanish, and even central moments satisfy the relation $M_{2n}= (2n-1)!![M_2]^n$. One can easily verify that $\overline{Z^{3}}^{\mu_0}\neq 0$ and 
$\overline{Z^{4}}^{\mu_0}\neq 3[\overline{Z^{2}}^{\mu_0}]^2$. 

{\it Introduction of boundaries.\/} For an infinite-medium problem, the asymmetry in the distribution can be taken into account by calculating the higher moments, and using the results to find a more realistic analytical form \cite{CLAX:05,Pury:90}. However, this strategy cannot be used when boundaries are present. Under such circumstances, GAP becomes, I suggest, worthy of consideration, because a Gaussian of the form
\begin{equation}
F(z,t)={1\over \sqrt{2\pi \sigma^2(t)}}\exp\left[-{\bigl \{z-m(t)\bigr \}^2\over 2 \sigma^2(t)}\right ],\\
\end{equation} 
satisfies the partial differential equation
\begin{equation}
\partial_t F(z,t)=\textstyle{1\over 2}a(t)\partial_{zz}F(z,t)-b(t)\partial_{z}F(z,t), 
\end{equation}
with $a=d\sigma^2/dt$ and  $b=dm/dt$ \cite{OrnsWijk:33}. Of course, there still remains the question of inferring the boundary conditions to be imposed on a solution of Eq.~(12); for further details, the reader is referred to a recent article \cite{RaziSigm:05}.

{\it The definition of $D$.\/} For defining $D$, it will be helpful to enlarge the scope of our discussion and recall the corresponding expressions for a Brownian particle (B-particle) of mass $m$ \cite{UhleOrns:31}:
\begin{eqnarray}
\overline{v}^{v_0}&=&v_0\exp(-\beta t),\\
\overline{z}^{v_0}&=&z_0+v_0(1-e^{-\beta t})/\beta,\\
\overline{v^2}^{v_0}&=&v_0^2e^{-2\beta t} + (kT/m) (1-e^{-2\beta t}),\\
\overline{Z^2}^{v_0}\!&=&(kT/f)\!\left [2 t-(3-4e^{-\beta t} + e^{-2\beta t})/\beta \right ],\\
\overline{Z^2\,} 
&=& (kT/f)\left[2t-2(1-e^{-\beta t})/\beta\right ].
\end{eqnarray} 
Here $v_0\equiv v(0)$, $v$ is the $z$-component of the velocity $(-\infty \leq v \leq \infty)$, $f$ denotes the friction coefficient, $\beta=f/m$, and the other symbols have their usual meanings \cite{UhleOrns:31}. U\&O obtained these results through the Langevin equation, but one can also utilize the Klein-Kramers equation (KKE)\cite{Chandra:43} and apply the procedure outlined above, choosing $\chi=z^m v^n$ and replacing integration over all $\mu$ by integration over all $v$; whatever the route, one finds that the density $F(z,t; z_0,v_0)$ remains a Gaussian at all times.  

The transient terms in Eqs.~(5)--(8) and (13)--(17) represent, it cannot be overemphasized, the ballistic phase, during which the velocity relaxes exponentially to its equilibrium value; diffusion, as envisaged by Einstein, starts {\it after\/} this relaxation is over.

CLAX have opted for the following definition:
\begin{equation}
\int_{-\infty}^{\infty}\!\!\!dz Z^2 F(z,t; z_0, \mu_0)= 2Dt.
\end{equation} 
To be exact, they write $2D_{zz}ct$ on the right-hand side, but I have absorbed the speed $c$ into the definition of $D$ so that its dimensions come out to be $[{\rm L}]^{2}[{\rm T}]^{-1}$. I submit that their definition is unreasonable because it implies that $D$ depends (not only on $t$ but also) on the initial conditions. The truth of this assertion, though implicit in the definition, will now be made manifest through an explicit calculation of the variance for three cases.

Thus far we have been occupied with particles with the same initial velocity ($\mu_0=1$). The CLAX expression for $D$ emerges upon dividing the right-hand side of Eq.~(10) by $2t$. Let us now consider two other situations: isotropic distribution of initial velocities ($\overline{\mu_0}=0, \overline{\mu_0^2}=1/3$) and a `magic-angle' incidence ($\overline{\mu_0}=1/\sqrt{3}, \overline{\mu_0^2}=1/3$). In the first case, Eq.~(6) gives $\overline{z}=z_0$, so that Eq.~(8) implies
\begin{equation}
\overline{Z^2\,} \equiv
\overline{z^2\,}- 
(\overline{z\,})^2
= \textstyle{1\over 3}c^2\tau\left[2t-2(1-e^{-\alpha t})/\alpha\right ],
\end{equation}
and in the second
\begin{equation}
\overline{Z^2\,}^{\mu_0}=\textstyle{1\over 3}c^2\tau\left[2t-(3-4e^{-\alpha t}+e^{-2\alpha t})/\alpha\right ].
\end{equation}
The definition advocated by CLAX would lead to a new expression of $D$ in each case. We will see later that the standard definition amounts to replacing the left-hand side of Eq.~(18) by its long-time limit ($t\to \infty$).

{\it A caution concerning the TE.\/} If one sets $\textstyle{1\over 3}c^2\tau=D=kT/f$ and $\alpha = \beta$, Eq.~(17) coincides with Eq.~(19), and with the result obtained by using the TE; this is to be expected because the TE can be obtained from the LBE or the KKE only if one assumes that the velocity distribution corresponds to equilibrium. The shortcomings of the TE were well exposed, within the context of the KKE, first by Hemmer \cite {Hemm:61} and later by Wilemski \cite{Wile:76}. Those who place their trust in the TE will do well to devote some attention to these penetrating analyses, which are not vitiated by a change in the transport equation.

{\it Some other special cases.\/} The formal identity of Eqs.~(16) and (20) is noteworthy, and so is the fact (which can be easily verified) that if one sets $g_2=0$ in the CLAX result for the variance, it coincides with  Eq.~(20). Unfortunately, this coincidence does not extend to the higher moments, which means that photons cannot be made to mimic Brownian motion by a judicious adjustment of the experimental conditions.

{\it Does $D$ change in an absorbing medium?\/}  If the absorption probability per unit time is a constant ($k$, say), the LBE will take the following form (with $\gamma \equiv\alpha+k$):
\begin{equation}
\Bigl [\partial_t +c\mu \partial_z \Bigr ]\Psi(z,\mu, t)= -\gamma\Psi(z,\mu,t)-\textstyle{1\over 2}\alpha F(z,t).
\end{equation}
When absorption is present, the densities  $\Psi(z,\mu, t)$ and $F(z,t)$ cannot be normalized. At first sight, this can be easily remedied, since the substitutions
\begin{equation}
\tilde {\Psi}(z,\mu,t)\equiv e^{kt}\Psi(z,\mu,t), \quad  \tilde {F}(z,t)\equiv e^{kt}F(z,t)\\
\end{equation} 
undo the loss caused by absorption. The claim made by CLAX, concerning the insensitivity of $D$ to $k$ is based essentially on this transformation. Though $\tilde {\Psi}(z,\mu,t)$ and $\tilde F(z,t)$ are properly normalized, their introduction is not sufficient to clinch the argument. Consider, for example, a system where absorption dominates. In this case, a particle will be absorbed at its first (or, at most, the second or the third) collision, and this is a time domain where ballistic behavior dominates the dynamics. I also add in passing that the above argument applies equally well to the KKE (or any other linear kinetic equation).

When a B-particle moves through the suspending medium, the direction of its velocity does not change much after a collision, and many collisions are needed to randomize an initially imparted velocity. In photon (or neutron) transport, sometimes called inverse Brownian motion \cite{Razi:94}, a particle with no or negligible mass collides with stationary targets, and ${\bf \Omega}$ is randomized after each collision (when scattering is isotropic). Nonetheless, if certain restrictions  are satisfied, the DE will apply to Brownian motion as well as its inverse \cite {Razi:94}. That the fundamental solution of the DE is a Gaussian \cite{Eins:05} is merely an affirmation of the central limit theorem, according to which the probability density function of the sum of $N$ independent, identically distributed random variables can be approximated, if $N$ is sufficiently large, by a Gaussian, regardless of how the individual variables are distributed \cite {Pury:90}. In other words, one need not worry about the nature of the particle and the host medium; conversely, the question of what is diffusing in what cannot be answered within the confines of the DE. Though Einstein was concerned essentially with B-particles, this preamble makes it clear that his derivation would remain valid, when suitably applied, to any system.
 
Let us therefore follow Einstein \cite{Eins:05}, and assume the existence of a time interval, say $\varepsilon$, which is very small compared to the time interval, say $\Upsilon$, over which observations are made, but sufficiently long to justify the assumption that the displacements suffered by the particle in two successive intervals, each of duration $\varepsilon$, can be viewed as mutually independent. Whence follows the equation
\begin{equation}
F(z,t+\varepsilon ) = \int_{-\infty}^{\infty}F(z-s,t)\phi_0(s,\varepsilon) ds,
\end{equation}
in which $\phi_0(s,\varepsilon)ds$ is the probability that a particle will go from $z-s$ to $z$ in a time interval $\varepsilon$. By virtue of the central limit theorem \cite{Pury:90}, we can ascribe a Gaussian form to $\phi_0(s,\varepsilon)$ with zero mean and a variance proportional to $\varepsilon$. Next, we expand $F(z-s,t)$ in a Taylor series, introduce the notation
$
\overline{s^k}= \int_{-\infty}^{\infty}s^k\phi_0(s,\varepsilon) ds,
$
and get
\begin{equation}
F(z,t+\varepsilon ) = F(z,t)+ \textstyle{1\over 2}\overline{s^2}\partial_{zz}F,
\end{equation}
which can be immediately transformed to the DE, $\partial_tF=D\partial_{zz}F$, with
\begin{equation}
D=\lim_{\varepsilon\to 0}{\overline{s^2}\over 2\varepsilon}.
\end{equation}
The limit $\varepsilon\to 0$ is purely formal, since we have already agreed that $\varepsilon \gg \tau$ (or $\alpha \varepsilon \gg 1$). In fact, it proves advantageous to replace the limit $\varepsilon\to 0$ with $\varepsilon\to \infty$, for $D$ then turns out to be the integral ($t=0$ to $t=\infty$) of the velocity autocorrelation function \cite{Kubo:68}; one gets $D=c^2\tau/3$ (photons) and $D=kT/f$ (B-particles).

Let us now apply the same reasoning to an absorbing medium (or to a non-absorbing medium but with a diffusing particle that is subject to first-order decay), and see if we can infer a  modified DE by replacing $\phi_0(s,\varepsilon)$ with $\phi_k(s,\varepsilon)$, the new transition probability. To make any progress, it seems necessary to assume that $\phi_k=\phi_0e^{-k\varepsilon}$, and then we obtain
\begin{equation}
F(z,t+\varepsilon )-F(z,t) = \Bigl [ -(1-e^{-k\varepsilon})+ \textstyle{1\over 2}\overline{s^2}e^{-k\varepsilon}\partial_{zz}\Bigr ]F(z,t) .
\end{equation}
On dividing by $\varepsilon$ and letting $\varepsilon\to 0$, we formally get
\begin{equation}
\partial_tF(z,t)=-kF(z,t)+ D\partial_{zz}F(z,t),
\end{equation}
but the limit is to be interpreted with some sensitivity, since we cannot transgress the condition $\alpha \varepsilon \gg 1$. The difficulty can be avoided---and, as we shall see, meaningful experiments made possible---if we impose the demand $k \varepsilon \ll 1$ before taking the limit. We conclude, then, that Eq.~(27), with the customary definition of $D$ \cite{Kubo:68}, is valid only if $k\ll\alpha$ (i.e., negligible absorption).

I elaborate on the above argument by using an analogy. Time-resolved photo-induced anisotropy measurements, which monitor the {\it rotational\/} diffusion of an electronically excited probe, are interpreted through the rotational DE. Though the definition of anisotropy incorporates a normalization akin to that in Eq.~(22), the experiment becomes inconclusive, owing to a precipitous decline in the signal-to-noise ratio, if $k\gg {\cal D}_r$, where the rotational diffusion coefficient ${\cal D}_r$ is the analog of $\alpha$. The lifetime of the probe $1/k$ plays the role of $\Upsilon$, effectively setting a scale for the observation time; the {\it only\/} remedy, when diffusion is slow, is to use a long-lived probe \cite{RaziNat:73, AustChanJovi:79}.

{\it Time-dependent and stationary versions of the DE.\/} Aronson and Corngold \cite {AronCorn:99} argue that $D$ depends on $k$, and base their claim on analyses of stationary versions of the LBE. However, these analyses lead not to $D$ as such, but to a quantity, called the diffusion length $L$, which is a measure of the average distance a particle would travel, when released in an infinite medium, before it is captured. Evidently, $L$ would  decrease with increasing absorption, and if one chooses to define $D$ through the relation $D=kL^2$, $D$ too will depend on $k$.  But a $D$ so inferred would pertain, if absorption is high, to a particle that is captured before it has taken a large number of diffusive steps; in the extreme case $k\gg \alpha$ the particle would be absorbed during its ballistic phase, and such a definition would be liable to an objection similar to that leveled against a time-dependent $D$, since the final form of the Poisson type equation depends also on the distribution of the sources \cite{Davi:57}. ``It is also necessary to remark,''  Davison \cite{Davi:57} informed his readers and I go along with him, ``that, though the value of $c$ [our $\alpha/\gamma$] in the system does not enter directly the criteria of applicability of the diffusion approximation, in practice the dimensions of the system will usually be of the order of $L$ \ldots, and this effectively limits the application of the theory to systems where $|1-c|$ is small''. 

The time-independent DE corresponds to a stationary (or steady) state, a situation when a constant rate of production of the particles within a region equals their rate of absorption. When this condition is fulfilled, the time-dependent DE, $\partial_tF(z,t)=D\partial_{zz}F(z,t)-kF(z,t)+S(z)$, where $S(z)$ denotes the source term, goes into the stationary version $D\partial_{zz}F_s(z)-k F_s(z)=-S(z)$, which leads, upon division throughout by $D$, to a Poisson type equation, formally identical with that deducible from the stationary LBE:  $\partial_{zz}F_s(z)- F_s(z)/L_0^2=-s(z)$, where $L_0^2=D/k$ and $s(z)=S(z)/D$. Smoluchowski was the first to realize that if a particle is produced at a random point in a medium containing traps, the trapping probability per unit time is, in general, a function of $t$ \cite{Chandra:43,Smol:16}. In other words, absorption or trapping differs from first-order decay (such as electronic de-excitation or nuclear disintegration). Aronson and Corngold \cite {AronCorn:99} did not take this complication into account when they wrote:``It is obvious (and is also well known) that the solution of the time-independent diffusion equation is obtained from that of the time-dependent equation by integrating over all time.'' It is far from straightforward---unless absorption is small---to relate the parameter $k$ appearing in the stationary DE to the corresponding time-dependent parameter which should appear in the time-dependent DE. The reader is referred to two reviews \cite{Razi:94, Szab:89} and a recent article \cite{RaziJorgEuri:04}.

{\it Conclusions.\/} The principal conclusions of this study will now  be stated:-- ($a$) Since the DE implies and is implied by the central limit theorem, an unequivocal definition of $D$ is possible only if the diffusing particle can execute a large number of diffusive steps; when this condition is satisfied, $D$ becomes independent of time $t$ and absorption probability $k$; if this restriction is ignored, $D$ will depend not only on $t$ and $k$, but also on the experimental conditions.  ($b$) The Gaussian approximation implies that photon transport can be described by  a non-Markovian partial differential equation. ($c$) Inferences drawn from the TE are suspect. ($d$) Smoluchowski's observation that the absorption probability per unit time depends on $t$ should be taken into account when the DE is compared with its stationary version.

\end{document}